\documentclass[aps,twocolumn,prd,nofootinbib,showpacs]{revtex4}
\usepackage[pdftex,bookmarks]{hyperref}

\usepackage[T1]{fontenc}
\usepackage{amsmath,amssymb}
\usepackage{graphics,wrapfig,times}
\usepackage{graphicx}
\usepackage{color}
\usepackage{pause}
\usepackage{mathrsfs}

 \makeatletter
 \newcommand{\pback}[1]{{
   \let\@rrow=\leftarrowfill
   \mathchoice{\AIN@stemPullBack{#1}{\@rrow}}{\AIN@stemPullBack{#1}{\@rrow}}
     {\AIN@indxPullBack{#1}{\@rrow}}{\AIN@indxPullBack{#1}{\@rrow}}}
   \vphantom{#1}}

 \newcommand{\AIN@stemPullBack}[2]{
   \vtop{\mathsurround=0pt
   \ialign{##\crcr$\textstyle{#1}\strut$\crcr
     \noalign{\kern-0.4ex\nointerlineskip}{\tiny#2}\crcr}}}

 \newcommand{\AIN@indxPullBack}[2]{
   \vtop{\mathsurround=0pt
   \ialign{##\crcr\hfil$\scriptstyle{#1}$\hfil\crcr
     \noalign{\kern+0.4ex\nointerlineskip}{\tiny#2}\crcr}}}

\def\bar{\overline}
\def\be{\begin{equation}}
\def\ee{\end{equation}}
\def\bea{\begin{eqnarray}}
\def\eea{\end{eqnarray}}
\def\ba{\begin{array}}
\def\ea{\end{array}}
\def\nn{\nonumber}
\def\w{\wedge}

\def\F{\mathcal{F}}
\def\G{\mathcal{G}}

\def\a{\alpha}
\def\b{\beta}
\def\c{\mu}
\def\d{\textrm{d}}

\definecolor{D}{rgb}{0.00,0.17,0.48}
\definecolor{M}{rgb}{0.00,0.02,0.83}
\definecolor{L}{rgb}{0.58,0.79,1.00}
\definecolor{R}{rgb}{0.80,0.00,0.00}
\definecolor{G}{rgb}{0.02,0.40,0.10}

\begin{document}

\title{Domain wall space-times with a cosmological constant}

\author{Chih-Hung Wang$^{1, 2, 3}$\footnote{Electronic address: chwang1101@phys.sinica.edu.tw}, Hing-Tong Cho$^{1}$\footnote{Electronic address: htcho@mail.tku.edu.tw}, and Yu-Huei Wu$^{3, 4}$\footnote{Electronic address: yhwu@mail.phy.ncu.edu.tw}}

\affiliation{1. Department of Physics, Tamkang University, Tamsui, Taipei 251, Taiwan.\\
 2. Institute of Physics, Academia Sinica, Taipei 115, Taiwan.\\
3. Department of Physics, National Central University, Chungli 320, Taiwan.\\
4. Center for Mathematics and Theoretical Physics, National Central University, Chungli 320, Taiwan.}

\begin{abstract}

We solve vacuum Einstein's field equations with the cosmological constant in space-times admitting 3-parameter group of isometries with 2-dimensional space-like orbits. The general exact solutions, which are represented in the advanced and retarded null coordinates, have two arbitrary functions due to the freedom of choosing null coordinates. In the thin-wall approximation, the Israel's junction conditions yield one constraint equation on these two functions in spherical, planar, and hyperbolic domain wall space-times with reflection symmetry. The remain freedom of choosing coordinates are completely fixed by requiring that when surface energy density $\sigma_0$ of domain walls vanishes, the metric solutions will return to some well-known solutions. It leads us to find a planar domain wall solution, which is conformally flat, in the de Sitter universe.

\end{abstract}
\date{\today}
\maketitle


\section{Introduction}\label{1}

Inflationary Universe was originally proposed to solve horizon and flatness problems in the hot big-bang cosmological model \cite{Guth-81}. From the discovery of the cosmic microwave background (CMB) anisotropy, the idea of cosmological inflation becomes more convincible and it serves as the initial conditions for the subsequent hot big bang. Most of popular inflation models are described by scalar fields, called inflaton fields, with their effective potentials \cite{LL-00}. Although there still lacks a fundamental theory to explain the origin of inflation,  some effective theories reveal that inflation may naturally happen due to the spontaneous symmetry breaking and phase transitions in the early Universe \cite{Linde-90}.

When a phase transition occurred in the early Universe, various types of topological defects, which are classified by homotopy groups, can form in the vacuum manifold and this phenomenon is known as the Kibble mechanism \cite{Kibble-76}. Therefore domain walls, which are a particular type of the topological defects, correspond to vacuumlike hypersurfaces interpolating between separate vacua. Beside the Kibble mechanism, domain walls can also form by quantum tunneling process of false vacuum decay, i.e. bubble nucleation \cite{Coleman-85}, or quantum production of topological defects in de Sitter space \cite{BGV-91} (see \cite{CS-97} for a review of domain walls). Ref. \cite{BGV-91} has showed that the topological defects can be continuously formed during inflation and still be present after inflation with appreciable densities.  Hence it motivates us to study the gravitational effects of domain walls in the de Sitter Universe.

Since we are only interested in the macroscopic effects of domain walls, it is sufficient to study the domain wall space-times in the thin-wall approximation, where the wall is regarded as an infinitely thin, with $\delta$-function singularity in the energy-momentum tensor. Therefore gravitational effects of domain walls are described by Einstein's field equations off the wall together with Israel's junction conditions \cite{Israel-66}. As far as we know,  domain wall solutions have been studied based on two different approaches. The first approach starts from exact solutions of Einstein's field equations off the wall in the specific coordinates, and then the wall's motion in the same coordinates is described by Israel junction conditions.  The second approach is to introduce the co-moving coordinates, where the wall is placed at a particular constant coordinate variable, say $z=0$, and then the exact solutions of Einstein's field equations off the wall are obtained in the co-moving coordinates.

Cveti\v{c} {\it et al}  \cite{CGS-93} studied the local and global properties of domain wall space-time with the cosmological constant $\Lambda$ in the co-moving coordinates. They found domain wall solutions based on three assumptions. The first one assumed that the two-dimensional spatial sections $V_2$ of space-times "parallel" to the wall are homogeneous and isotropic. It corresponds to space-times admitting 3-parameter group of isometries with 2-dimensional space-like orbits, i.e. $V_2$ are 2-dimensional spheres, planes, or hyperboloids \cite{KSHM-80, GS-70}. The second one assumed that the space-time section orthogonal to the wall, say $(t, z)$-plane, is static. It means that the metric components $g_{tt}$ and $g_{zz}$ are $t$-independent. The third assumption required that the directions parallel to the wall are boost invariant, i.e. extrinsic curvature of constant-$z$ hypersurfaces is boost invariant. It yields that the metric function intrinsic to $V_2$ are separable. It seems to us that the second and third assumptions are not satisfactory since they may eliminate some interesting domain wall solutions. For example, our planar domain wall solutions in the de Sitter Universe, which are obtained in Sec. \ref{4-2} (see (\ref{g-6})), is conformally flat and will return to the metric of the de Sitter Universe when surface energy density of the domain wall vanishes. The solution (\ref{g-6}) cannot be found in \cite{CGS-93} since it does not satisfy the second and third assumptions.

In this paper, we release the last two assumptions in \cite{CGS-93} and find the general solutions of Einstein's field equations with $\Lambda$ in the double-null coordinates. Solving Einstein's equations in null coordinates has been used to find Schwartschild and Reissner-Nordstr\"om solutions in \cite{Chandrasekhar-83}. Since the non-degenerate general solutions contain two arbitrary functions $F(v)$ and $G(u)$ due to the freedom of the null-coordinate choices, the Israel's junction conditions yield one constraint equation on these two functions, where the domain wall is placed at a constant-$z$ hypersurface.

From the generalized Birkhoff theorem \cite{KSHM-80, GS-70},  one can find a coordinate transformation (see Eq. (\ref{T-R})) to make our non-degenerate solutions become
\bea
g= -U(R)\, \d T \otimes \d T + \frac{1}{U(R)} \d R \otimes \d R + R^2  \d V_2, \label{static-g}
\eea where $U(R)=K- 2M/R - (\Lambda/3) R^2$ and $\d V_2 = (1-K r^2)^{-1} \d r \otimes \d r +r^2 \d \phi \otimes \d \phi$. It is clear that the metric (\ref{static-g}) is static in the certain range of coordinates. Moreover, it turns out that the domain wall, which is originally sitting at a constant-$z$ hypersurface, becomes moving in the coordinates $(T, R, r, \phi)$. So one may expect that a domain wall solution in the co-moving coordinates is locally equivalent to a moving wall embedding in a spacetime with metric (\ref{static-g}).\footnote{It is possible that the solutions obtained in these two coordinate systems may have different global structure of spacetime.} Here, we call these two different approaches the comoving-coordiante approach and the moving-wall approach.  The equivalence of these two approaches has been demonstrated by Bowcock \textit{et al} \cite{BCG-00, thank}.\footnote{In \cite{BCG-00}, these two different approaches are called the brane-based approach and the bulk-based approach.}, so there exists coordinate transformations between these two approaches. Having shown this equivalence Bowcock {\it et al} solved Israel's junction conditions to find the most general brane-universe solutions based on the moving-wall approach. In the study of brane cosmologies, it is more suitable to use the moving-wall approach.

In this paper, we solve Israel's junction conditions and find the general domain-wall solutions in the comoving-coordinate approach.  We will show that the domain-wall solutions in the comoving-coordinates approach are more useful to study gravitational perturbations and quantum fluctuations than moving-wall approach, which is normally adopted to study the dynamics of brane-universe. It is known that a proper coordinate choice can largely simplify physical problems and equations. For example, when one studies metric perturbations in a specific background spacetime, a proper choice of background coordinates is important since it may largely simplify perturbed equations or make these equations solvable. Moreover, when we study quantum fluctuations in curved spacetime, the coordinate choices become significant since there is no coordinate-invariant definition of the vacuum state, i.e. the vacuum state is observer dependent \cite{BD-82}. In the comoving-coordinate approach, the Israel's junction conditions only fix one freedom of double-null coordinate choices, so the remain freedom can be used to further simplify our domain-wall solutions or to avoid coordinate singularities appeared in metric solutions.  However, in the moving-wall approach, the meric solutions outside the wall are fixed in the static form, i.e. Eq. (\ref{static-g}), which have coordinate singularities at some finite values of $R$, and the Israel's junction conditions will completely determined the trajectories of walls. In this work, we propose a reasonable way to fix the remain coordinate freedom by requiring that when surface energy density $\sigma_0$ of domain walls vanishes, the metric solutions will return to some well-known solutions. For example, when $\sigma_0=0$,  planar domain wall solutions with $M=0$ and $\Lambda>0$ (see (\ref{g-6})) will return to de Sitter Universe in conformal-time coordinates, where the quantum fluctuations have been largely studied. So the domain-wall effects on primordial quantum fluctuations in early Universe may clearly be seen by studying quantum fluctuations in the solution (\ref{g-6}).

The plan of this paper is as follows. In Sec. \ref{2}, we briefly review the thin-wall approximation and Israel's formalism in the covariant approach. Sec. \ref{3} presents general solutions of Einstein's field equations with $\Lambda$ in space-times admitting 3-parameter group of isometries with 2-dimensional space-like orbits. The general non-degenerate solutions have $F(v)$, $G(u)$ and three parameters: 2-dimensional constant curvature $K$, gravitational mass $M$, and $\Lambda$. We also show that the solutions will return to some well-known exact solutions in specific coordinates by choosing $F$ and $G$. In Sec. \ref{4}, we limit our discussion in space-time being reflection symmetry with respect to the wall, and the Israel's junction conditions require $F$ and $G$ satisfying an algebraic equation. Sec. \ref{4-1} discusses spherical and hyperbolic domain walls with $M=0$ in the de Sitter Universe. In Sec. \ref{4-2}, we  present a planar domain wall solution, which is conformally flat, in the de Sitter Universe. When surface energy of the domain wall vanishes, the solution returns to de Sitter metric in the conformal time.

We use the units $\hbar=c=1$, and the metric signature is $(- + + +)$. The Latin indices $a, b, \cdots$ are referred to coordinate indices and the Greek indices $\a,\b,\gamma \cdots$ referred to orthonormal frame indices. $g$ and $\nabla$ denote metric tensor and Levi-Civita connection, respectively.


\section{The thin-wall approximation} \label{2}

In the thin-wall approximation, the thickness $\varepsilon$ of a thin wall is taken to be zero, so the infinitely thin wall becomes a 3-dimensional timelike, null or spacelike hypersurface $\Sigma$ in  4-dimensional space-times, and its associated stress-energy tensor $T^a{_b}$ of the space-times has a $\delta$-function singularity on $\Sigma$. Here, we will assume $\Sigma$ to be a 3-dimensional timelike hypersurface for our current interest.  From Einstein's field equations and the singular property of $T^a{_b}$, it turns out that the extrinsic curvature $\pi_{ab}$ of $\Sigma$ has a jump discontinuity across $\Sigma$ and its discontinuity is naturally related to surface stress-energy distribution of matter fields on $\Sigma$ \cite{Israel-66, Ipser-Sikivie-84}.

\subsection{Israel's Junction conditions}

Since $\Sigma$ is a 3-dimensional timelike hypersurface, one may first introduce its unit spacelike normal $n$ satisfying $g(n, n)= 1$, and the intrinsic metric $h$ of $\Sigma$ is then given by
\bea
h= g - \tilde{n} \otimes \tilde{n}, \nn
\eea where $\tilde{n}= g(n, -)$ is the metric dual of $n$. So the extrinsic curvature $\pi_{ab}$ of $\Sigma$ is defined by
\bea
\pi_{ab}  = \frac{1}{2}(\mathcal{L}_n \bar{h} )_{ab}|_\Sigma, \label{pi}
\eea where $\mathcal{L}_n$ denotes the Lie derivative along the unit normal $n$, and $\bar{h}$ is any extension of $h$ to a neighborhood of $\Sigma$. Because of the discontinuity of $\pi_{ab}$ across $\Sigma$, it is convenient to introduce the notation $\pi_{ab}|_{\pm}$, where the subscripts $\pm$ refer to values just off the surface on the side determined by the direction of $\pm n$. In \cite{Israel-66}, it showed that
\bea
\gamma_{ab}\equiv \pi_{ab}|_+ - \pi_{ab}|_- = - \kappa (S_{ab} - \frac{1}{2} h_{ab} S_c{^c}) \label{IJC}
\eea in the thin-wall limit $\varepsilon \rightarrow 0$, where $\kappa= 8 \pi G_N$ and
%
$S_{ab}\equiv \lim_{\varepsilon \rightarrow 0}\int_0^{\varepsilon} \d l\, T_{ab}$ denotes surface stress-energy tensor. Here $l$ is the proper distance through $\Sigma$ in the direction of $n$. Eq. (\ref{IJC}) is normally called Israel's junction conditions. It is worth to point out that Eq. (\ref{IJC}) is also valid when we include the cosmological constant $\Lambda$ in Einstein's field equations. One may further impose the space-time geometry being reflection symmetric with respect to $\Sigma$, which will be considered in Sec. \ref{4}, and it yields
%
$\pi_{ab}|_+ = - \pi_{ab}|_-$.

\subsection{The surface stress-energy tensor}

The surface stress-energy tensor $S= S_{ab}\, \d x^a \otimes \d x^b$ is usually assumed to have the following perfect-fluid form
\bea
S= (\sigma - \tau ) \, \tilde{u} \otimes \tilde{u} - \tau\, h,
\eea where $\sigma$ and $\tau$ denotes surface energy density and tension of $\Sigma$, respectively. $\tilde{u}= h(u, -)$ is the intrinsic metric dual of the unit timelike vector field $u$, which lies within $\Sigma$.
It is known that dynamics of $S$ can be determined by Eq. (\ref{IJC}), Einstein's field equations and the Gauss-Codazzi equations in the thin-wall approximation \cite{Israel-66, Ipser-Sikivie-84}. In particular, one of the Gauss-Codazzi equation yields the conservation equation of $S$, which is
\bea
D \cdot S = (\sigma- \tau) D_u \tilde{u} + \tilde{u} \, D \cdot [(\sigma- \tau) u] - \d \tau =0, \label{conservation-S}
\eea where $D$ is the 3-dimensional intrinsic covariant derivative on $\Sigma$ satisfying $D h =0$ and torsion-free condition. $D \,\cdot$ denotes the divergence. In terms of coordinate components, it is easy to show that, for any tensor field $T$ on $\Sigma$,
\bea
D_a T^{b...c}{_{d...e}} = h_a{^p}h_q{^b}\cdots h_r{^c}h_d{^s}\cdots h_e{^t} \nabla_p \bar{T}{^{q...r}}{_{s...t}},
\eea where $\bar{T}$ is any extension of $T$ to a neighborhood of $\Sigma$.

For dust walls, i.e. $\tau=0$, Eq. (\ref{conservation-S}) yields that the world lines of $u$ are geodesics and the surface energy density is conserved, which is $D \cdot (\sigma u)=0$.  For domain walls, i.e. $\tau=\sigma$, we simply obtain $\d \sigma=0$, which means that $\sigma=\sigma_0$ is a constant on $\Sigma$.
In the following discussion (see Sec. \ref{4}), we will only concentrate on domain wall space-times with reflection symmetry, so Eq. (\ref{IJC}) becomes
\bea
\gamma_{ab}= 2 \pi_{ab}|_+ = - 2 \pi_{ab}|_- = -\frac{\kappa \sigma_0}{2} \,h_{ab}. \label{IJC-1}
\eea Therefore the reflection symmetric domain-wall space-time will be described by vacuum solutions of Einstein's field equations with the cosmological constant off $\Sigma$ and Eq. (\ref{IJC-1}).


\section{Space-times admitting 3-parameter group of isometries} \label{3}

It is still a great challenge for mathematicians and physicists to find an exact solution of Einstein's field equations without assuming any symmetry of space-time. Most of the well-known exact solutions are found in spaces of high symmetry, so the group of isometries becomes a useful method to classify and also to find exact solutions \cite{KSHM-80}. In Sec. \ref{4}, we will consider that the domain wall is homogeneous and isotropic in its two space dimensions, so it leads us to assume that the 4-dimensional space-time geometry induced by the domain wall source has the same symmetric property, i.e. 2-dimensional spatial sections $V_2$ parallel to the wall are homogeneous and isotropic \cite{CGS-93}. This assumption of space-time symmetry may correspond to space-times admitting 3-parameter group of isometries with 2-dimensional spacelike orbits, denoted by $G_3(2, s)$. In this section we shall solve the vacuum Einstein's field equations with cosmological constant off $\Sigma$ under $G_3(2, s)$.

It is known that a $n$-dimensional Riemannian space $V_n$ admitting $G_q$, where $q={n(n+1)/2}$, is a space of constant curvature \cite{KSHM-80}. Hence the orbits $V_2$ of $G_3(2, s)$ must have constant Gaussian curvature $K$, and correspond to 2-dimensional spheres $(K>0)$, planes $(K=0)$, or hyperboloids $(K<0)$.  Moreover, it is always possible to introduce coordinates $(t, z, r, \phi)$ such that the metric tensor $g$ with $G_3(2, s)$ has the form \cite{Goenner-70, GS-70}
\bea
g= e^{2 \nu(t, z)} (- \d t \otimes \d t + \d z \otimes \d z ) + e^{2 \lambda(t, z)} \d V_2, \label{g}
\eea where $\d V_2 = (1-K r^2)^{-1} \d r \otimes \d r +r^2 \d \phi \otimes \d \phi$. By rescaling $e^{2\lambda}$, one can normalize the constant curvature to be $K = +1, 0 , -1$. It is clear that the two space dimensions of $\Sigma$ is placed at $t, z$=constant. In \cite{CGS-93}, its metric ansatz yields
\bea
g= e^{2\nu(z)} (- \d t \otimes \d t + \d z \otimes \d z ) + e^{2 [\alpha(t)+\beta(z)]} \d V_2,
\eea which is only a special form of the metric (\ref{g}). We shall stress again that the general exact solutions of $\nu(t, z)$ and $\lambda(t, z)$ obtained in this section will only be valid off $\Sigma$.

We first introduce the Einstein's 3-forms \cite{BT-87}
\bea
G_\c = R_{\a\b}\w * (e^\a \w e^\b \w e_\c),
\eea where $R_{\a\b}$ are curvature 2-forms defined in terms of Levi-Civita connection $\nabla$, $e^\a$ are orthonormal co-frames, and $*$ denotes the Hodge map associated with $g$. So Einstein's equations with $\Lambda$ are
\bea
G_\c =- 2 \kappa\, \tau_\c +  2 \Lambda * e_\c, \label{Einstein-eq}
\eea where $\tau_\c$ are stress-energy 3-forms of matter fields. Since we are only interested in vacuum solutions of Eq. (\ref{Einstein-eq}),  $\tau_\c$ will be assumed to vanish in the following calculation. By substituting the metric (\ref{g}) into vacuum Eq. (\ref{Einstein-eq}) yields
\bea
\dot{\lambda} \,\lambda' - \dot{\nu}\, \lambda' + {\dot{\lambda}}' - \nu' \dot{\lambda} =0, \label{G_0}\\
 e^{-2\nu} ({\dot{\lambda}}^2 - 3 {\lambda'}^{2} - 2 \lambda'' + 2 \dot{\lambda}\dot{\nu} + 2 \lambda' \nu' ) + e^{-2\lambda} K = \Lambda,\\
 e^{-2\nu}(3 {\dot{\lambda}}^2 - {\lambda'}^2 + 2 \ddot{\lambda} - 2 \dot{\lambda}\dot{\nu} - 2 \lambda' \nu') + e^{-2\lambda} K = \Lambda, \\
 e^{-2\nu} (\ddot{\nu} - \nu'' + \ddot{\lambda} - \lambda'' + {\dot{\lambda}}^2 - {\lambda'}^2 ) = \Lambda \label{G_1},
\eea where dots and primes here and in the following denote the differentiation with respect to $t$ and $z$, respectively.  These four non-linear partial differential equations, Eqs. (\ref{G_0})-(\ref{G_1}), are difficult to find an analytic nontrivial solution due to their highly coupling. However, they can be largely simplified by introducing advanced and retarded null coordinates $u$, $v$ defined by $u=\frac{1}{2} (t+z)$, $v=\frac{1}{2}(t-z)$. A similar procedure has been used to find Schwartzschild and Reissner-Nordstr\"om solutions and their maximally analytic extensions. \cite{Chandrasekhar-83}.

It is convenient to set
\bea
A(t, z) = e^{2\nu}  , \hspace{0.5cm} B^2(t, z)= e^{2\lambda} ,
\eea where $A >0$, so the metric (\ref{g}) becomes
\bea
g= A(t, z)  (- \d t \otimes \d t + \d z \otimes \d z ) + B^2 (t, z) \d V_2. \label{g-0-1}
\eea
By transforming $(t, z)$ coordinate variables to the null coordinates $(u, v)$, Eqs. (\ref{G_0})-(\ref{G_1}) with some linear combination yield
\bea
&&A B_{,uu} - A_{,u} B_{,u} =0, \label{chandra-1}\\
&&A B_{,vv} - A_{,v} B_{,v} =0, \label{chandra-2}\\
&&B B_{,uv} + B_{,u} B_{,v} + A K = \Lambda A B^2, \label{chandra-3}\\
&& B\, (\ln A)_{,uv} +2\, B_{,uv} =2\, \Lambda AB \label{chandra-4},
\eea where a subscript comma denotes partial differentiation with respect to the coordinates following it. It turns out that Eqs.  (\ref{chandra-1})-(\ref{chandra-4})  become much simpler and solvable. If we put $K= 1$ and $\Lambda =0$, Eqs. (\ref{chandra-1})-(\ref{chandra-4}) agree with Eqs. (31)-(34) in Sec. 17 of \cite{Chandrasekhar-83}\footnote{we shall point out the metric signature $(+, -, -, -)$ used in \cite{Chandrasekhar-83} is different from ours.}.

From Eqs. (\ref{chandra-1}) and (\ref{chandra-2}), we observe that there exists degenerate non-trivial solutions in the case of $B_{,u}$ or $B_{,v}$ vanishing. Hence it is better to study the degenerate and non-degenerate solutions separately. In the special case of $K=\Lambda=0$, i.e. plane symmetry, the degenerate and non-degenerate solutions have been studied in \cite{Taub-51, Ipser-Sikivie-84}, where \cite{Ipser-Sikivie-84} called them class-I and class-II solutions, respectively.

\subsection{The degenerate case: $B_{,u}=0$ or $B_{,v}=0$ (but not both)}

Since Eqs. (\ref{chandra-1})-(\ref{chandra-4}) are invaraint under switching the coordiante variables $u$ and $v$, it is only necessary to study either $B_{,u}=0$ or $B_{,v}=0$. Suppose $B_{,u}=0$ and $B_{,v}\neq 0$, i.e. $B=B(v)$. Then Eq. (\ref{chandra-1}) are trivial satisfying and Eq. (\ref{chandra-3}) yields
\bea
B^2 = \frac{K}{\Lambda} = \textrm{const.} \geqslant 0, \label{chandra-5-1}
\eea where $K, \Lambda \neq 0$. It is clear that both $B_{,u}$ and $B_{,v}$ vanish, which gives a trivial solution. So the case of $K, \Lambda \neq 0$ does not give a degenerate non-trivial solution, and we will not proceed our discussion in this case further.

For $K=0$, i.e. plane symmetry, Eq. (\ref{chandra-3}) yields $\Lambda=0$ since $A$ and $B$ cannot vanish. It means that the non-trivial degenerate solutions with plane symmetry cannot allow the non-vanishing cosmological constant.  Then we expect to recover the class-I solutions in \cite{Ipser-Sikivie-84}. One may first solve Eq. (\ref{chandra-2}) to yield
$A= G(u) B_{,v},$ where $G$ is an arbitrary function of $u$, and the Eq. (\ref{chandra-4}) becomes trivial satisfying. So this solution does return to class-I solutions of \cite{Ipser-Sikivie-84}. It is worth to mention that the planar domain-wall solutions obtained in \cite{Ipser-Sikivie-84, Vilenkin-83} have the class-I solutions outside the wall.

\subsection{The non-degenerate case: $B_{,u}\neq 0$ and $B_{,v}\neq 0$}

In this case, Eq. (\ref{chandra-1}) and Eq. (\ref{chandra-2}) yield
\bea
&& A(u, v)= F(v)\,B_{,u}, \label{chandra-6} \\
&& A(u, v)= G(u)\, B_{,v}, \label{chandra-7}
\eea respectively, where $F(v)\neq 0$ and $G(u) \neq 0$ are arbitrary functions of their arguments. Substituting Eq. (\ref{chandra-6}) into Eq. (\ref{chandra-3}) gives
\bea
(B B_{,v}+ K F(v)\, B - \frac{\Lambda}{3} F(v)\, B^3 )_{,u}=0. \label{chandra-8}
\eea Hence, from Eq. (\ref{chandra-8}), we obtain
\bea
B_{,v} = - K F(v) + \frac{H(v)}{B} + \frac{\Lambda}{3} F(v) B^2, \label{chandra-9}
\eea where $H(v)$ is an arbitrary function of $v$. A similar result can be obtained by substituting Eq. (\ref{chandra-7}) into Eq. (\ref{chandra-3}), which yields
\bea
B_{,u} = - K G(u) + \frac{J(u)}{B} + \frac{\Lambda}{3} G(u) B^2, \label{chandra-10}
\eea where $J(u)$ is an arbitrary function of $u$. Since Eqs. (\ref{chandra-9})-(\ref{chandra-10}) possess some symmetry between $u$ and $v$, we multiply Eqs. (\ref{chandra-9}) and (\ref{chandra-10}), and use Eqs.(\ref{chandra-6})-(\ref{chandra-7}), which gives
\bea
B_{,u} B_{,v} &=& -K A + \frac{J(u)}{B} B_{,v} + \frac{\Lambda}{3} A B^2 \label{chandra-11}\\
&=&-K A + \frac{H(v)}{B} B_{,u} + \frac{\Lambda}{3} A B^2. \label{chandra-12}
\eea If $\{J(u), H(v)\} \neq 0$, Eqs. (\ref{chandra-11}) and (\ref{chandra-12}) yield
\bea
\frac{J(u)}{H(v)} = \frac{B_{,u}}{B_{,v}}= \frac{G(u)}{F(v)}.
\eea So we finally obtain
\bea
\frac{J(u)}{G(u)}= \frac{H(v)}{F(v)}= 2 M, \label{chandra-13}
\eea where $M$ is a constant. By substituting Eqs. (\ref{chandra-6}), (\ref{chandra-10}) into Eq. (\ref{chandra-4}) and using Eqs. (\ref{chandra-3}) and (\ref{chandra-13}), one can easily verify that Eq. (\ref{chandra-4}) is automatically satisfying.
Therefore the general exact solution of Eqs. (\ref{chandra-1})-(\ref{chandra-4}) yields
\bea
A = - F(v) G(u) \left( K  - \frac{2 M}{B} - \frac{\Lambda}{3} B^2 \right) \label{chandra-14}
\eea with $B(u, v)$ satisfying
\bea
\d B= - \left( K  - \frac{2 M}{B} - \frac{\Lambda}{3} B^2 \right) (G(u) \,\d u + F(v) \,\d v). \label{chandra-15}
\eea Also, the metric Eq. (\ref{g-0-1}) becomes
\bea
g= 4 F(v) G(u) ( K  - \frac{2 M}{B} - \frac{\Lambda}{3} B^2 ) \d u \otimes \d v + B^2 \d V_2. \label{g-1}
\eea
From the metric (\ref{g-1}), it is clear to see that the existence of the two arbitrary functions $F(v)$ and $G(u)$ is due to the freedom of choosing null coordinates $u, v$. Hence the different choices of $F(v)$ and $G(u)$ may correspond to the metric $g$ in different null coordinates. According to the generalized Birkhoff theorem \cite{KSHM-80, GS-70}, the metric (\ref{g-1}) can actually be put into a static form (see Eq. (\ref{static-g})) by introducing coordinates
\bea
T= \int F(v) \d v + \int G(u) \d u, \hspace{0.3cm} R = B. \label{T-R}
\eea In the following, we show that the metric  (\ref{g-1}) will return to some well-known solutions in various different coordinates by choosing three parameters $K$, $\Lambda$, $M$, and two functions $F(v)$, $G(u)$.

In the case of $K=1$ and $\Lambda=0$, Eqs. (\ref{chandra-14}) and (\ref{chandra-15}) yields the Schwartzschild solution obtained in \cite{Chandrasekhar-83}, and $M$ will be regarded as gravitational mass. Moreover, by setting
\bea
F(v) = \frac{2M}{v} \hspace{0.5cm} \textrm{and} \hspace{0.5cm} G(u)=\frac{2M}{u}, \label{FG}
\eea
we obtain the Schwartzschild solution in the Kruskal coordinates, which is the maximal extension of the Schwartzschild solution \cite{Chandrasekhar-83, HE-73}. If one further set $v/u= e^{t/2M}$, the metric (\ref{g-1}) will be transited to Schwartzschild coordinates $(t, R, \theta, \phi)$, which yields
\bea
g= - h(R) \d t \otimes \d t + h(R)^{-1} \d R \otimes \d R + R^2 \,\d \Omega_2,
\eea  where $B\equiv R$ and $h(R)= 1- 2M/R$. $\d \Omega_2 \equiv \d \theta \otimes \d \theta + \sin^2 \theta \,\d \phi \otimes \d \phi$ denotes surface element of 2-spheres. It is clear that the coordinates $(t, R)$ only cover the regions of both $u, v$ being positive and negative in the $(u, v)$-plane.

In the case of $K=\Lambda=0$, i.e. plane symmetry with vanishing cosmological constant, it is not difficult to verify that Eqs. (\ref{chandra-14}) and (\ref{chandra-15}) give the class-II solutions of \cite{Ipser-Sikivie-84}.\footnote{It should be noticed that the two arbitrary functions $F$ and $G$ in \cite{Ipser-Sikivie-84} are different from ours.}


In the case of $K=1$, $M=0$ and $\Lambda > 0$,
by setting
\bea
F(v)=  \sqrt{\frac{3}{4 \Lambda}}\,\frac{1}{v}\hspace{0.5cm} \textrm{and} \hspace{0.5cm} G(u)= \sqrt{\frac{3}{4 \Lambda}}\,\frac{1}{u}, \label{FG-1}
\eea
we obtain the de Sitter metric in the Kruskal coordinates:
\bea
g= - \frac{12}{\Lambda} \frac{1}{(uv -1)^2} \,\d u \otimes \d v + \frac{3}{\Lambda} \frac{(uv+1)^2}{(uv-1)^2}\, \d \Omega_2. \label{g-kruskal}
\eea  Its global properties have been discussed in \cite{GH-77}.

Finally, we consider $K=M=0$, which will be useful for cosmological models. Since $K=M=0$, Eq. (\ref{chandra-15}) gives
\bea
- \frac{3}{\Lambda} \frac{1}{B} + c =  \int G(u) \d u + \int F(v) \d v, \label{B}
\eea where $c$ is a constant of integration, and the metric (\ref{g-1}) becomes
\bea
g= B^2 \left(- \frac{4 \Lambda}{3} F(v) G(u) \,\d u \otimes \d v + \d X_2 \right), \label{g-2}
\eea where $\d X_2 = \d x \otimes \d x + \d y \otimes \d y$. It is interesting to discuss the sign of $F$ and $G$ in the cases of $\Lambda$ being positive or negative.  For $\Lambda>0$, the metric (\ref{g-2})
indicates that $FG >0$. On the other hand, the metric (\ref{g-2}) yields $F G <0$ in the case of $\Lambda<0$. Since we are interested in the de Sitter Universe, we will only consider $\Lambda>0$.

It is clear that the metric (\ref{g-2}) is conformally flat if we simply choose
\bea
F(v) = - \sqrt{\frac{3}{ \Lambda}}, \hspace{0.2cm} \hspace{0.2cm} G(u) =  - \sqrt{\frac{3}{ \Lambda}}, \label{FG-2}
\eea for $\Lambda >0$, and Eq. (\ref{B}) yields
\bea
B= \sqrt{\frac{3}{ \Lambda}}\, \frac{1}{(u + v - c)}.
\eea
%
Furthermore, the metric (\ref{g-2}) becomes
%
\bea
g= \frac{1}{\frac{\Lambda}{3} \eta^2}\, (-\d \eta \otimes \d \eta + \d {z} \otimes \d {z} + \d {X}_2), \label{g-4}
\eea where $\eta= t + c$. The metric (\ref{g-4}) is a well-known solution of a flat expanding de Sitter Universe in the conformal time \cite{Mukhanov-05}. One can also realize that the metric (\ref{g-4}) is a background metric for describing a slow-roll inflation in the early Universe \cite{LL-00}.

It is known that the metric solution (\ref{g-1}) is only valid off the $\Sigma$. Hence a general solution of domain wall space-time should also need to satisfy the Israel's junction conditions. As we mentioned in Sec. \ref{1}, the study of the domain wall space-times can be separated into two different approaches mainly due to the different choices of coordinates, namely moving-wall approach and comoving-coordiante approach. The moving-wall approach is usually used to study the two space-times, $M_+$ and $M_-$ with a common moving boundary $\Sigma$ \cite{GV-89, FHS-07, BCG-00}. In this approach, the metric solutions of $M_+$ and $M_-$ have the static form (\ref{static-g}), and it turns out that $\Sigma$ is moving in this coordinate system. Therefore the Israel's Junction conditions become equations of motion for $\Sigma$.

The comoving-coordinate approach is to introduce the co-moving coordinates of the wall system, i.e. the rest frame of the wall $\Sigma$ \cite{CGS-93}.  In the co-moving coordinates, the wall $\Sigma$ is normally placed at a constant $z$-coordinate position, say $z=0$, so Israel's Junction conditions serve as boundary conditions of the metric solutions at $z=0$.  In Sec. \ref{4}, we will adopt the second approach by considering a domain wall sitting at $z=0$ with space-time being reflection symmetry. It turns out that the Israel's junction condition, i.e. Eq. (\ref{IJC-1}), yields some constraints on $F(v)$ and $G(u)$ at $z=0$, so $F(v)$ and $G(u)$ cannot be arbitrarily chosen.

\section{Domain wall space-times} \label{4}

In this section, we limit our discussion on reflection-symmetric domain wall space-times, so Israel's junction conditions yield Eq. (\ref{IJC-1}) and constant surface energy density $\sigma=\tau= \sigma_0$.
Since the metric solution (\ref{g-1}) has a freedom of choosing double null coordinates $(u, v)$, we may simply assume that $\Sigma$ is placed at $z \equiv  u - v =0$. According to the reflection symmetry, it only needs to study the metric (\ref{g-1}) at $z>0$. Hence Eq. (\ref{IJC-1}) may be considered as the boundary conditions of the metric (\ref{g-1}) at $z=0$.

Since $\Sigma$ is placed at $z=0$, the unit normal $n$ and unit time-like vector $u$ in the coordinates $(t, z, r, \phi)$ become
\bea
n=\frac{1}{ \sqrt{A(t, z)}} \,\,\partial_z, \hspace{1cm} u=\frac{1}{ \sqrt{A(t, z)}}\, \,\partial_t,
\eea where $\partial_t$ and $\partial_z$ denote the coordinate basis, and the intrinsic metric $h$ of $\Sigma$ yields
\bea
h= - A|_{z=0}\, \d t \otimes \d t + B^2|_{z=0}\,\d V_2. \label{h}
\eea  By substituting (\ref{h}) into Eq. (\ref{IJC-1}) and using Eq. (\ref{pi}) gives two non-vanishing equations
%
%
\bea
&&A^{\prime}|_+= - \frac{\kappa\sigma_0}{2}A^{\frac{3}{2}}|_{z=0}, \label{A}\\
 &&B^{\prime}|_+ = - \frac{\kappa\sigma_0}{4} \sqrt{A} B|_{z=0}. \label{B'}
\eea It is clear that Eqs. (\ref{A}) and (\ref{B'}) will give $F(v)$ and $G(u)$ further constraints, so $F(v)$ and $G(u)$ cannot be arbitrary functions.

By substituting the solution (\ref{chandra-14}) into Eqs. (\ref{A}) and (\ref{B'}) and using (\ref{chandra-15}), a tedious but straightforward calculation yields
\begin{widetext}
\bea
\left.\left(-\frac{F_{,v}}{F} + \frac{G_{,u}}{G} + \frac{L_{,u}- L_{,v}}{L}\right)\right|_{z=0} &=& \left. -\kappa \sigma_0 \sqrt{- F G L}\,\right|_{z=0},\label{A-1} \\
 \left. L (F - G)\right|_{z=0}&=&\left. - \frac{\kappa\sigma_0}{2} ( B\,\sqrt{- F G L}\,)\right|_{z=0}, \label{B'-1}
\eea
\end{widetext} where
\bea
L(u, v) := K  - \frac{2 M}{B} - \frac{\Lambda}{3} B^2.
\eea However, Eqs. (\ref{A-1}) and (\ref{B'-1}) actually are not independent. To verify this, one may first differentiate Eq. (\ref{B'-1}) with respect to $t$.
Then by using Eqs. (\ref{chandra-15}) and (\ref{B'-1}), one can show that Eq. (\ref{A-1}) is implied by Eq. (\ref{B'-1}) if $( F+G )|_{z=0} \neq 0$, which is assumed in the following discussion. It is interesting to notice that when $(F - G)|_{z=0}=0$, Eq. (\ref{B'-1}) yields $\sigma_0=0$. We then obtain an important result that if the domain wall exists, i.e. $\sigma_0 \neq 0$, in the space-time, one cannot choose $F(t)=G(t)$. It turns out that Eqs. (\ref{FG}), (\ref{FG-1}) and (\ref{FG-2}) are not valid choices in the domain wall space-time. We shall mention that Eq. (\ref{B'-1}) cannot uniquely determine the $F(v)$ and $G(u)$, so one still has freedom of choosing these two functions.


\subsection{Spherical and hyperbolic domain walls} \label{4-1}

In this subsection, we study spherical and hyperbolic domain walls in the de Sitter Universe, i.e. $K=\pm 1$, $M=0$, and $\Lambda>0$. It is convenient to introduce
\bea
\F(v)= \int F(v) \d v, \hspace{0.5cm} \G(u) = \int G(u) \d u,
\eea and in the case of $K=1$, Eq. (\ref{chandra-15}) yields
\bea
B= \left\{ \begin{array}{l} - \sqrt{\frac{3}{\Lambda}} \coth\left[\sqrt{\frac{\Lambda}{3}}(\F + \G)\right] \hspace{0.2cm} \textrm{for}\hspace{0.2cm}B^2> \frac{3}{\Lambda}, \vspace{0.2cm}\label{B'-2}\\
- \sqrt{\frac{3}{\Lambda}} \tanh\left[\sqrt{\frac{\Lambda}{3}}(\F + \G)\right] \hspace{0.4cm} \textrm{for}\hspace{0.2cm}B^2< \frac{3}{\Lambda}. \end{array} \right.
\eea  It is known that $B^2= \frac{3}{\Lambda}$ corresponds to cosmological horizons \cite{GH-77}. Substituting Eq. (\ref{B'-2}) into Eq. (\ref{B'-1}) gives
\bea
\begin{array}{l}\frac{\dot{f}_-}{\sqrt{\dot{f}^2_+ - \dot{f}^2_-}} =  \mp \frac{\kappa\sigma_0 }{4}\sqrt{\frac{3}{\Lambda}} \cosh\left[\sqrt{\frac{\Lambda}{3}}\,f_+\right] \hspace{0.4cm} \textrm{for}\hspace{0.2cm}B^2> \frac{3}{\Lambda}, \vspace{0.2cm}\label{B'-2-1} \\
\frac{\dot{f}_-}{\sqrt{\dot{f}^2_- - \dot{f}^2_+}} =  -  \frac{\kappa\sigma_0}{4} \sqrt{\frac{3}{\Lambda}} \sinh\left[\sqrt{\frac{\Lambda}{3}}\,f_+\right]
 \hspace{0.2cm} \textrm{for}\hspace{0.2cm}B^2< \frac{3}{\Lambda},
 \end{array}
\eea where $f_+(t) := (\F + \G)|_{z=0}$ and $f_- (t):= (\F - \G)|_{z=0}$.  It is clear that $\dot{f}_\pm = \frac{1}{2}(F \pm G)|_{z=0}$. The $\mp$ sign corresponds to $f_+>0$ or $f_+<0$, respectively. Eq. (\ref{B'-2-1}) indicates that either $f_+$ or $f_-$ is given, the other will be determined.

In the case of $K=-1$, a similar calculation yields
\be\ba{l}
 B =  \sqrt{\frac{3}{\Lambda}} \tan \left[\sqrt{\frac{\Lambda}{3}}(\F + \G)\right],
\ea\ee and
\be\ba{l}
\frac{\dot{f}_-}{\sqrt{\dot{f}^2_+ - \dot{f}^2_-}} =  \frac{\kappa\sigma_0}{4}\sqrt{\frac{3}{\Lambda}}  \sin\left[\sqrt{\frac{\Lambda}{3}}\,f_+\right],
\ea\ee for $\sec\left[\sqrt{\frac{\Lambda}{3}}(\F + \G)\right]>0$. We expect that the choices of $f_+$ or $f_-$ are related to global properties of space-times and a proper choice of $f_+$ or $f_-$ may yield a simpler domain wall solution and also avoid coordinate singularities. So far, our guiding principle of choosing $f_+$ or $f_-$ is that when $\sigma_0=0$, the domain wall solutions should return to some well-known solutions. Here, we present an example of choosing $f_+$ in the case of $K=1$. A more comprehensive study on $f_+$, $f_-$ and also global properties of domain-wall space-times will be present in our following work.

In the case of $K=1$, the de Sitter metric (\ref{g-kruskal}) in Kruskal coordinates has been presented in Sec. \ref{3}. From Eq. (\ref{FG-1}), we know that $f_+ = \frac{1}{2}\sqrt{\frac{3}{\Lambda}}\ln t$ for $t>0$, and then substituting $f_+$ into Eq. (\ref{B'-2-1}) yields
\bea
\dot{f}_-= \frac{- \sqrt{\frac{3}{\Lambda}}\,\,(t+ 1)}{2\,t\,\,\sqrt{ \frac{\Lambda}{3} (\frac{8}{ \kappa \sigma_0})^2\,\, t + (t + 1)^2 }},
\eea where we only consider $B^2>\frac{3}{\Lambda}$. We then obtain $F$ and $G$, which are
\be\ba{l}
F(v) = \sqrt{\frac{3}{4 \Lambda}}\frac{1}{v}\left( 1 - \frac{(v+ 1)}{\,\sqrt{ \frac{\Lambda}{3} (\frac{8}{ \kappa \sigma_0})^2\, v + (v + 1)^2 }} \right), \label{fg}\\
G(u) = \sqrt{\frac{3}{4 \Lambda}}\frac{1}{u}\left( 1 + \frac{(u+ 1)}{\,\sqrt{ \frac{\Lambda}{3} (\frac{8}{ \kappa \sigma_0})^2\, u + (u + 1)^2 }} \right).
\ea\ee It is clear that when $\sigma_0=0$, Eq. (\ref{fg}) returns to Eq. (\ref{FG-1}), and the metric should become de Sitter metric in Kruskal coordinates.

\subsection{Planar domain walls} \label{4-2}

In this subsection, we study planar domain wall in the de Sitter Universe, i.e. $K=M=0$ and $\Lambda>0$. In the planar domain wall case, the Israel junction condition (\ref{B'-1}) becomes
\be\ba{l}
\frac{\dot{f}_-}{\sqrt{\dot{f}^2_+ - \dot{f}^2_-}} =  \pm \frac{\kappa\sigma_0 }{4}\sqrt{\frac{3}{\Lambda}}, \label{B'-2-3}
\ea\ee which is much simpler. The $\pm$ sign corresponds to $B>0$ or $B<0$, respectively. In Sec. \ref{3}, we have found the metric (\ref{g-4}), which is conformally flat. It was obtained by choosing $F=G=\textrm{constant}$. We observe that if $F$ and $G$ are two different constants, Eq. (\ref{B'-2-3}) can also be satisfied.
Therefore we consider
\bea
F(v) = F_0, \hspace{0.2cm} \hspace{0.2cm} G(u) =  F_0 \,\Gamma \label{FG-3}
\eea where $F_0$ and $\Gamma$ are constants. Since $FG>0$ for $\Lambda>0$, so $\Gamma>0$. If one considers $F_0 <0$, then by substituting Eq. (\ref{FG-3}) into (\ref{B'-1}) yields
\bea
\sqrt{\frac{\Lambda}{3}}\, (1- \Gamma) = \mp \frac{\kappa\sigma_0}{2} \,\sqrt{\Gamma}, \label{B'-3}
\eea for $B$ being positive or negative. If one further assumes $\sigma_0 \geqslant 0$, which is physically reasonable,  Eq. (\ref{B'-3}) yields that $\Gamma \geqslant 1$ for $B>0$, and $\Gamma\leqslant 1$ for $B<0$.  The algebraic equation (\ref{B'-3}) has two real positive roots, which are
\bea
\Gamma = 1 + \frac{3 \kappa^2 \sigma_0^2}{8 \Lambda}\pm \frac{\sqrt{48 \kappa^2 \sigma_0^2 \,\Lambda + 9 \kappa^4 \sigma_0^4}}{8 \Lambda}, \label{Gamma}
\eea so  the larger root of $\Gamma$ corresponds to the case of $B>0$, and the smaller root for $B<0$.

Substituting Eq. (\ref{FG-3}) into Eqs. (\ref{B}) and (\ref{g-2}) yields
\bea
&&B= - \frac{3}{\Lambda F_0\, (\,\Gamma u + v - c_0)},\\
&&g= B^2 \left(\frac{\Lambda\Gamma}{3} F_0^2 (-\d t \otimes \d t + \d z \otimes \d z) + \d X_2\right).
\eea It is interesting to notice that $B$ should take different values of $\Gamma$ for $B>0$ and $B<0$. So these two domain regions may need to be considered separately. Since $F_0$ is an arbitrary negative constant, we may choose $F_0 =- \sqrt{\frac{3}{\Lambda\Gamma}}$, and then obtain
\bea
g= \frac{-\d \eta \otimes \d \eta + \d {z} \otimes \d {z} + \d {X}_2}{\frac{\Lambda\Gamma}{3} \left(\frac{\Gamma+1}{2}\,\eta + \frac{\Gamma-1}{2} \,z\right)^2}\label{g-5}
\eea for $z>0$. It is obvious that when $\Gamma=1$, i.e. $\sigma_0=0$, the metric (\ref{g-5}) returns to (\ref{g-4}), which is the de Sitter Universe. Moreover, the metric (\ref{g-5}) is conformally flat, though it has $z$-dependence. According to the reflection symmetry, we then obtain the domain-wall solution in de Sitter Universe:
\bea
g= \frac{-\d \eta \otimes \d \eta + \d {z} \otimes \d {z} + \d {X}_2}{\frac{\Lambda\Gamma(\Gamma+1)^2}{12} \left(\,\eta + \frac{\Gamma-1}{\Gamma+1}\, |z| \right)^2}. \label{g-6}
\eea
The metric solution (\ref{g-6}) is important and useful for cosmologists to study the domain-wall effects in the early Universe. It is worth to mention that the metric (\ref{g-6}) will not return to the planar domain-wall solution obtained in \cite{Ipser-Sikivie-84, Vilenkin-83} when $\Lambda=0$, since the metric (\ref{g-6}) outside the domain wall is belong to the non-degenerate solution. Moreover, $\Lambda=0$ will make the metric (\ref{g-6}) become divergent. Finally, we would like point out that if one apply the following coordinate transformations:
\be\ba{l}
\tilde{\eta}= \frac{\Gamma + 1}{2} \eta + \frac{\Gamma - 1}{2} z,\hspace{0.3cm} \tilde{z}= \frac{\Gamma - 1}{2} \eta + \frac{\Gamma + 1}{2} z, \\
\tilde{x}=\sqrt{\Gamma} x, \hspace{1.9cm}\tilde{y}= \sqrt{\Gamma} y,
\ea\ee for $z>0$, the metric (\ref{g-5}) also returns to (\ref{g-4}). However, in this coordinates ($\tilde{\eta}, \tilde{z}, \tilde{x}, \tilde{y}$), the wall's location becomes
\be\ba{l}
\tilde{z}= (\frac{\Gamma-1}{\Gamma+1}) \tilde{\eta},
\ea\ee which is moving along $\tilde{z}$.

Here, we only consider the spherical, planar, and hyperbolic domain-wall space-time with $M=0$ and $\Lambda>0$. It might also be interesting to study more general cases, e.g. $M \neq 0$ and $\Lambda<0$. Moreover, the global properties of these domain wall solutions are important and will be discuss in our next work. In particular, the problems of the two-bubbles collision in the early Universe exhibit the hyperbolic symmetry on 2-dimensional spatial section $V_2$ of space-times  \cite{HMS-82, FHS-07}, so it should be important to further investigate on the global properties of $K= -1$.





\section{Conclusions}

We have systematically studied exact solutions of vacuum Einstein's field equations with $\Lambda$ in space-times admitting $G_3(2, s)$. The general solutions are classified into degenerate and non-degenerate solutions. In the degenerate case of $\{K, \Lambda\}\neq0$, we did not find any physically interesting solutions. However,  in the case of either $K=0$ or $\Lambda=0$, the degenerate field equations required the other to be vanishing. Therefore we obtained the planar symmetric solutions with vanishing $\Lambda$, which are equivalent to the class-I solutions of \cite{Ipser-Sikivie-84}. In the non-degenerate case, the general exact solutions are obtained in the double null coordinates, and they contain three parameters $K$, $M$, $\Lambda$, and two arbitrary functions $F(v)$ and $G(u)$ due to the freedom of choosing null coordinates. We considered non-degenerate vacuum solutions as the domain-wall solutions outside the wall.

In the domain-wall space-times, we assumed that the domain wall $\Sigma$ is homogeneous and isotropic in its two space dimensions. Since the non-degenerate solutions have freedom on choosing null coordinates, we then considered that $\Sigma$ is placed at $z=0$, and the Israel's junction conditions yield one constraint equation on $F(v)$ and $G(u)$. The remain freedom of choosing $F(v)$ and $G(u)$ is fixed by requiring that when surface energy density $\sigma_0$ of domain walls vanishes, the metric solutions will return to some well-known solutions. In the derivation of Israel's junction conditions, we only assume the space-time being reflection-symmetric without putting any restriction on the three parameters. Applying the Israel's junction conditions to the case of $M=0$ and $\Lambda>0$, we first discuss the cases of $K=\pm 1$. It turns out that Israel's junction conditions become a first-order ordinary differential equation for two arbitrary functions $f_+(t)$ and $f_-(t)$. An example of choosing $f_+$ has been presented.
In the case of $K=0$ , we obtained the planar domain-wall solution, which is conformally flat, in the de Sitter Universe.

We plan to use the solution (\ref{g-6}) to study primordial quantum fluctuation during inflation in the early Universe. Since the wall is rest at $z=0$ and the planar domain-wall metric is conformally flat, the quantum fluctuation of scalar fields may be solved exactly. Moreover, the study of global properties of the domain wall solutions obtained in this paper is in progress.
Beside the study of the planar domain walls, it is also interesting to extend our current investigation
to $\{ M \neq 0, \,\Lambda<0\}$.
\acknowledgments

The authors would like to thank Prof Kin-Wang Ng and Dr I-Chin Wang for helpful discussions. We also thank to Institute of Physics, Academia Sinica, where the part of work was done there.  CHW was supported by the National Science Council of the Republic of China under the grants NSC 98-2811-M-032-005. HTC was supported in part by National Science Council of the Republic of China under the Grant NSC 99-2112-M-032-003-MY3, and the National Center for Theoretical Sciences.  YHW was supported by Center for Mathematics and Theoretical Physics, National Central University.


--------------------------


%





%

%


\begin{thebibliography}{10}
%

\bibitem{Guth-81} A. H. Guth,  Phys. Rev. {\bf D 23}, 347 (1981).

\bibitem{LL-00} A. R. Liddle and D. H. Lyth, {\it Cosmological Inflation and Large Scale Structure}  (Cambridge University Press, Cambridge, 2000).

\bibitem{Linde-90}
Linde A D, {\it Particle Physics and Inflationary Cosmology} (Harwood Academic Publishers, Chur, Switzerland, 1990)


\bibitem{Kibble-76} T. W. B. Kibble,  J. Phys. {\bf A9}, 1387 (1976).

\bibitem{Coleman-85} S. Coleman,  {\it Aspects of Symmetry} ( Cambridge University Press, Cambridge, England, 1985).

\bibitem{BGV-91} R. Basu, A. H. Guth, and A. Vilenkin,  Phys. Rev. {\bf D 44}, 340 (1991).


\bibitem{CS-97} M. Cveti\v{c} and H. H. Soleng, Phys. Rep. {\bf 282}, 159 (1997).



\bibitem{Israel-66} W. Israel, Nuovo Cimento {\bf 44B}, 1 (1966).

\bibitem{CGS-93} M. Cveti\v{c}, S. Griffies, and H. H. Soleng, Phys. Rev. {\bf D 48}, 2613 (1993).

\bibitem{KSHM-80} D. Kramer, H. Stephani, E. Herlt, and M. MacCallum, {\it Exact solutions of Einstein's Field Equations} (Cambridge University Press, Cambridge, 1980).

\bibitem{GS-70} H. Goenner  and J. Stachel, J. Math. Phys. {\bf 11}, 3358 (1970).




\bibitem{Chandrasekhar-83} S. Chandrasekhar, \textit{The Mathematical Theory of Black
Hole} (Oxford University Press, New York, 1983).

\bibitem{BCG-00} P. Bowcock, C. Charmousis, and R. Gregory, Class. Quantum Grav. {\bf 17}, 4745 (2000).

\bibitem{thank} We would like to thank the referee for introducing us the Bowcock {\it et al}'s work and point out the similar techniques used in this paper and theirs \cite{BCG-00}.


\bibitem{BD-82} N. D. Birrell and P. C. W. Davies, \textit{Quantum fields in curved space}, (Cambridge University Press, Cambridge, 1982).

\bibitem{Ipser-Sikivie-84} J. Ipser and P. Sikivie, Phys. Rev. {\bf D 30}, 712 (1984).

\bibitem{Goenner-70} H. Goenner, Commun. Math. Phys. {\bf 16}, 34 (1970).


\bibitem{BT-87} I. M. Benn  and R. W. Tucker, {\it An introduction to spinors and geometry with applications to physics} (Bristol: Institute of Physics Publishing 1987)

\bibitem{Taub-51} A. H. Taub, Ann. Math. {\bf 53}, 472 (1951).



\bibitem{Vilenkin-83} A. Vilenkin,  Phys. Lett.   {\bf B 136}  47 (1983).


\bibitem{HE-73} S. W. Hawking and G. F. R. Ellis,  \textit{The Large Scale Structure of Space-Time} (Cambridge University Press, Cambridge, England, 1973).

\bibitem{GH-77} G. W. Gibbons and S. W. Hawking,  Phys. Rev. {\bf D 15}, 2738 (1977).

\bibitem{Mukhanov-05} V. Mukhanov,  \textit{Physical Foundations of Cosmology} (Cambridge University Press, Cambridge, England, 2005).

\bibitem{GV-89} D. Garfinkle and C. Vuille, Class. Quantum Grav. \textbf{6}, 1819, (1989).

\bibitem{HMS-82} S. W. Hawking, I. G. Moss, and J. M. Stewart, Phys. Rev. {\bf D 26}, 2681 (1983).

\bibitem{FHS-07} B. Freivogel, G. T. Horowitz, and S. Shenker, J. High Energy Phys.  {\bf 05} (2007) 090.





%














\end{thebibliography}
\end{document}